\documentclass[5p,twocolumn,preprint,times]{elsarticle}
\journal{Control Engineering Practice}

\usepackage{myStyle}
\usepackage{subcaption}

\usepackage{eso-pic}

\usepackage[hidelinks]{hyperref}
\AddToShipoutPictureBG*{%
	\AtPageUpperLeft{%
		\setlength\unitlength{1in}%
		\hspace*{\dimexpr0.5\paperwidth\relax} 
		\makebox(0,-0.75)[c]{\begin{tabular}{c c}
				Max van Haren, Roy S. Smith and Tom Oomen, System Identification Beyond the Nyquist Frequency: A Kernel-Regularized Approach, \\
				Control Engineering Practice 164:106425 (2025), DOI: \href{https://doi.org/10.1016/j.conengprac.2025.106425}{10.1016/j.conengprac.2025.106425}, uploaded June 17th \\
		\end{tabular}}
}}

\renewcommand{\citep}{\cite}
\renewcommand{\citet}{\cite}

\newif\ifthesismode
\thesismodefalse  

\begin{document}
	\begin{frontmatter}
	\title{System Identification Beyond the Nyquist Frequency: A Kernel-Regularized Approach\tnoteref{label1}}
	\tnotetext[label1]{This research has received funding from the ECSEL Joint Undertaking under grant agreement 101007311 (IMOCO4.E), which receives support from the European Union Horizon 2020 research and innovation programme.}
	\author[1]{Max van Haren\corref{cor1}}
	\cortext[cor1]{Corresponding author.}
	\ead{m.j.v.haren@tue.nl}
	\author[2]{Roy S. Smith}
	\author[1,3]{Tom Oomen}
	\affiliation[1]{organization={Department of Mechanical Engineering, Control Systems Technology Section, Eindhoven University of Technology},
		addressline={Groene Loper 5},
		city={Eindhoven},
		postcode={5612 AE}, 
		country={The Netherlands}.}
	\affiliation[2]{organization={Automatic Control Laboratory, ETH Zürich},
		city={Zürich},
		postcode={8092}, 
		country={Switzerland}.}
	\affiliation[3]{organization={Delft Center for Systems and Control, Delft University of Technology},
		addressline={Mekelweg 2},
		city={Delft},
		postcode={2628 CN}, 
		country={The Netherlands}.}
	\begin{abstract}
		Models that contain intersample behavior are important for control design of systems with slow-rate outputs. The aim of this \manuscript is to develop a system identification technique for fast-rate models of systems where only slow-rate output measurements are available, e.g., vision-in-the-loop systems. In this \manuscript, the intersample response is estimated by identifying fast-rate models through least-squares criteria, and the limitations of these models are determined. In addition, a method is developed that surpasses these limitations and is capable of estimating unique fast-rate models of arbitrary order by regularizing the least-squares estimate. The developed method utilizes fast-rate inputs and slow-rate output \textcolor{reviewblue}{measurements} and identifies fast-rate models accurately in a single identification experiment. Finally, both simulation and experimental validation on a prototype wafer stage demonstrate the effectiveness of the framework.
	\end{abstract}
\end{frontmatter}
\section{Introduction}
Systems which have their outputs sampled at a reduced sampling rate compared to their inputs are becoming more prevalent, particularly in applications such as vision-in-the-loop systems \citep{Fujimoto2004}. Slow-rate output measurements generally constrain the identified models to reflect only slow system dynamics. However, fast-rate models are crucial for capturing higher-frequency behavior, especially beyond the Nyquist frequency. Therefore, fast-rate models of multirate systems are essential for intersample performance estimation \citep{Sivashankar1993, Chen1995} and sampled-data or multirate control \citep{Bamieh1991,Li2002,Oomen2007}.

System identification plays an essential role in understanding and modeling the dynamic behavior of systems for the use in control design. System identification finds broad application in areas such as digital twin creation \citep{Cech2022} and control design \citep{Schmidt2020}. Two distinct approaches are non-parametric and parametric system identification. Non-parametric system identification techniques, such as Frequency Response Functions (FRFs) \citep{Pintelon2012}, model systems without requiring a predefined structure. Unlike non-parametric approaches, parametric system identification generally involves a fixed model structure, e.g., using prediction error methods \citep{Soderstrom1989,Ljung1998}. A key development in system identification involves the application of kernel methods \citep{Ljung2020}. The unique properties of kernels offer a major advantage by enforcing high-level properties on the identified models, that in addition enable the identification of continuous-time models \citep{Pillonetto2014}. The high-level properties that are enforced by kernels include for example exponential decay and correlation of impulse response coefficients. For linear time-invariant equidistantly sampled systems, both parametric and non-parametric system identification techniques have been proven to work effectively.

Important developments have been made in identifying fast-rate models from slow-rate output measurements, primarily in terms of continuous-time and low-order parametric identification. First, continuous-time identification aims to identify a parametric continuous-time model using input-output data, as outlined in \citet{Unbehauen1990}. Typically, these methods require intersample assumptions on the input signals, e.g., zero-order hold or band-limited signals \citep{Ljung2009}, and hence, do not utilize broadband fast-rate inputs. Second, several methods are developed for fast-rate parametric model identification using slow-rate output measurements, including methods for identification of low-order Finite Impulse-Response (FIR) \citep{Ding2004} and output-error \citep{Zhu2009} models. Subspace \citep{Li2001} or hierarchical identification techniques \citep{Ding2011}, which identify a lifted system representation, are also developed. Generally, these methods identify a fast-rate model using slow-rate output measurements by incorporating prior information through a certain model structure, which is limited to low-order models and therefore does not include non-parametric models.

A key development in non-parametric identification for fast-rate models using slow-rate output measurements incorporates prior information through a local smoothness assumption on the frequency response \citep{VanHaren2023a}. The smoothness assumption allows one to disentangle aliased components in the frequency-domain. However, the local smoothness assumption is an approximation, since it assumes that the system can be approximated as a polynomial of a certain degree in the frequency domain, leading to a bias-variance trade-off \citep[Section~7.2.6]{Pintelon2012}.

Although methods to identify fast-rate models using slow-rate output measurements have been developed for certain scenarios, current approaches are limited to specific cases, including low-order parametric modeling and frequency-domain approaches. The aim of this \manuscript is to present a general framework to identify non-parametric fast-rate models using slow-rate output measurements, enabling more generic prior information. The developed method utilizes regularized estimators, which have been proven to work effectively in system identification for equidistantly sampled systems \citep{Ljung2020}. Specifically, system properties are enforced through the use of regularization or kernel techniques, expressed in either the frequency domain, similarly to \citet{Lataire2016} and \citet{Hallemans2022}, or the time domain, that is illustrated in this \manuscript. Compared to earlier regularized system identification approaches, the developed approach uniquely combines and utilizes slow-rate output measurements, fast-rate input measurements, and prior information of the system to identify fast-rate non-parametric models using slow-rate output measurements. The key contributions are the following.
\begin{itemize}
	\item[C1] Extending well-established conditions for unique model identification to the identification of fast-rate FIR models using slow-rate output measurements, which are critical for practical applications (\secRef{MRKernel:sec:reducedFIR}).
	\item[C2] Identification of arbitrary order fast-rate models using slow-rate output measurements, including non-parametric models, by incorporating prior information through kernel regularization (\secRef{MRKernel:sec:KernelFIR}).
	\item[C3] Validation of the framework for identification of fast-rate models for systems with slow-rate output measurements using a simulation and an experimental setup consisting of a prototype wafer stage (\secRef{MRKernel:sec:simulations} and \secRef{MRKernel:sec:exp}).
\end{itemize}

\paragraph*{Notation}
Fast-rate and slow-rate signals are denoted by subscript $h$ and $l$, with sampling times and frequencies $\tsh=\frac{1}{\fsh}$ and $\tsl=\fac\tsh=\frac{1}{\fsl}$, with downsampling factor $\fac\in\mathbb{N}_{1}$ and positive integers $\mathbb{N}$, discrete-time indices $\dt\in \mathbb{N}_{[0,N-1]}$ and $m\in \mathbb{N}_{[0,M-1]}$, with $N,M$ being the number of samples. Fast-rate and slow-rate discrete Fourier transforms are denoted with 
\begin{equation}
	\begin{aligned}
		X_h(e^{j\omega_k\tsh}) &=  \sum_{\dt=0}^{N-1}x_h(\dt)e^{-j\omega_k\tsh \dt}, \\
		X_l(e^{j\omega_k\tsl}) &= \sum_{m=0}^{M-1}x_l(m)e^{-j\omega_k\tsl m},
	\end{aligned}
\end{equation}
with complex variable $j=\sqrt{-1}$, frequency grid $\omega_k=2\pi\frac{k}{N}\fsh=2\pi\frac{k}{M}\fsl$, and frequency bins $k\in\mathbb{N}$. The fast-rate forward shift operator is denoted by $q$, i.e., $q^{-1}x_h(\dt)=x_h(\dt-1)$. The Kronecker product is denoted with $\otimes$.
\section{Problem Formulation}
In this section, the considered problem is presented. The identification setting and identifiability problem associated to fast-rate models using slow-rate output measurements are illustrated. Finally, the problem addressed in this \manuscript is defined.
\subsection{Identification Setting}
\begin{figure}[tb]
	\centering
	\vspace{3mm}\includegraphics{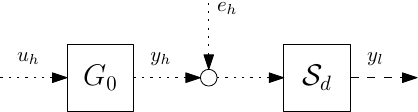}
	\makeatletter
	\caption{Identification setup for systems with slow-rate output measurements.}
	\label{MRKernel:fig:setting}
\end{figure}

The goal is to identify fast-rate model $G$ using slow-rate output measurements $y_l(m)=\mathcal{S}_dy_h(\dt)=y_h(m\fac)$ and fast-rate inputs $u_h$, having sampling rates $\fsl$ and $\fsh=\fac\fsl$ and known downsampler $\mathcal{S}_d$. The open-loop identification setting is visualized in \figRef{MRKernel:fig:setting}. 
The true fast-rate system $G_0$ is a linear time-invariant, single-input single-output system, and is described using the causal discrete-time state-space equations
\begin{equation}
	\begin{aligned}
		G_0:\:\: \begin{cases}
			{x}(\dt+1) &= Ax(\dt)+Bu_h(\dt), \\
			y_h(\dt) &= Cx(\dt) + Du_h(\dt).
		\end{cases}
	\end{aligned}
\end{equation}
The output is disturbed with zero-mean, independent, and identically distributed noise $e_h$, with variance $\sigma^2$, and downsampled into 
\begin{equation}
	\label{MRKernel:eq:DownsampleTimeDomain}
	y_l(m) = \mathcal{S}_d\left(y_h(\dt)+e_h(\dt)\right)=y_h(m\fac)+e_h(m\fac).
\end{equation}
\subsection{Identifiability of Fast-Rate Models}
In this section, identifiability of fast-rate models for systems with slow-rate outputs is investigated. The system $G_0$ is modeled using the FIR filter $G(q,\theta)$, which has fast-rate model output
\begin{equation}
	\label{MRKernel:eq:fastSampledOutputTime}
	\begin{aligned}
		\hat{y}_h(\dt) =  G(q,\theta)u_h(\dt)= \sum_{i=0}^{P-1} \theta_i q^{-i} u_h(\dt),
	\end{aligned}
\end{equation}
with model order $P\in \mathbb{N}_{[1,N]}$, and FRF 
\begin{equation}
\label{MRKernel:eq:ModelFRF}
	G(e^{j\omega\tsh})= \sum_{i=0}^{N-1}\theta_i e^{-j\omega\tsh i},
\end{equation}
which can be evaluated at any frequency $\omega\in\mathbb{R}$. Similarly to \eqref{MRKernel:eq:DownsampleTimeDomain}, the slow-rate model output is downsampled as $\hat{y}_l(m)=\mathcal{S}_d\hat{y}_h(\dt) = \hat{y}_h(m\fac)$. 

Identifiability of fast-rate models essentially involves determining whether the fast-rate input and the model structure allow for distinguishing between parameter values given the slow-rate output, which relates closely to the definition in \citet{Ljung1994}. Specifically, the fast-rate input should be sufficiently informative to distinguish between different parameter sets $\theta$. Additionally, the model structure $G(q,\theta)$ should allow for distinguishing between different parameter sets $\theta$, provided a sufficiently informative input. The identifiability of fast-rate models is defined in \defRef{def:Identifiability}.
\begin{definition}
	\label{def:Identifiability}
	The FIR coefficients $\theta$ of fast-rate model $G(q,\theta)$ in \eqref{MRKernel:eq:fastSampledOutputTime} are identifiable from slow-rate outputs if
	\begin{equation}
		\begin{aligned}
			\mathcal{S}_dG(q,\theta)u_h(\dt)=\mathcal{S}_dG(q,\theta^*)u_h(\dt), \quad \Rightarrow \theta=\theta^*. \\
		\end{aligned}
	\end{equation}
\end{definition}

Fast-rate models are not always identifiable from slow-rate outputs, where several examples are illustrated in \exampleRef{example:NonUnique}.
\begin{example}
	\label{example:NonUnique}
	Consider the model structure $G\left(q,\theta\right)$ \eqref{MRKernel:eq:fastSampledOutputTime} for three parameter values $\theta$, $\theta^*$, and $\theta^\prime$ with order $P=1000$. Let $u_h$ be a fast-rate impulsive input with $N=P=1000$ samples, being persistently exciting of order $P=N$ \citep[Lemma~13.1]{Ljung1998}. After downsampling by $\fac=4$, resulting in $M=250$ samples in the slow-rate output $y_l$, the responses of the three models are identical, i.e.,
	\[\mathcal{S}_dG(q,\theta)u_h(\dt)=\mathcal{S}_dG(q,\theta^*)u_h(\dt)=\mathcal{S}_dG(q,\theta^\prime)u_h(\dt),\]
	 as illustrated in the top of \figRef{MRKernel:fig:NonIdentifiableExample}. However, these are generated by different fast-rate models, as highlighted by the fast-rate FRFs and impulse response coefficients in the bottom of \figRef{MRKernel:fig:NonIdentifiableExample}, indicating that the model is not identifiable according to \defRef{def:Identifiability}.
\begin{figure}[tb]
	\centering
	\ifthesismode
	\includegraphics{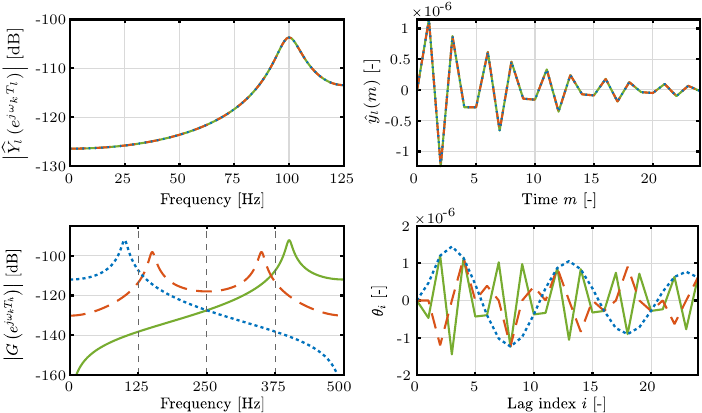}
	\else
		\includegraphics[width=\linewidth]{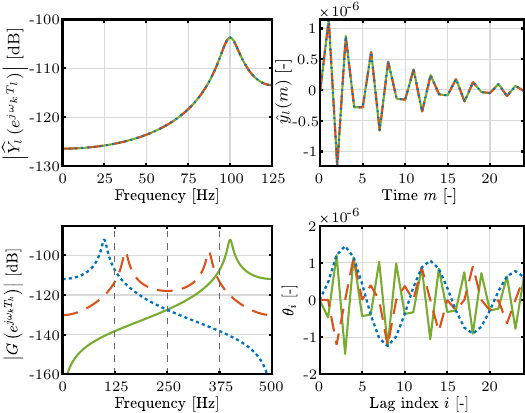}
	\fi
	\caption{A non-identifiable model structure $G(q,\theta)$ for three specific impulse response coefficients $\theta$ \markerline{mgreen}, $\theta^*$ \markerline{mred}[densely dashed], and $\theta^\prime$ \markerline{mblue}[densely dotted] (bottom right), shown by their equal downsampled ouput in time-domain $\hat{y}_l$ (top right) and frequency-domain $\widehat{Y}_l\left(e^{jk\omega_kT_l}\right)$ (top left) for impulsive input $u_h$.}
	\label{MRKernel:fig:NonIdentifiableExample}
\end{figure}
\end{example}

\subsection{Problem Definition and Approach}
\label{MRKernel:sec:problemDefApproach}
The fast-rate model $G(q,\theta)$ from \eqref{MRKernel:eq:fastSampledOutputTime} is not always identifiable from slow-rate outputs given an arbitrary model order $P$ and fast-rate inputs $u_h$, as illustrated by \exampleRef{example:NonUnique}.

The problem considered in this \manuscript is as follows. Given the fast-rate input signal $u_h$ containing $N$ samples and slow-rate output measurements $y_l$ containing $M$ samples, determine a fast-rate model $G(q,\theta)$ of arbitrary order  $P\in\mathbb{N}_{[1,N]}$, by incorporating appropriate prior information of the system. 
\section{Identification of Slow-Rate Systems}
\label{MRKernel:sec:reducedFIR}
\ifthesismode
	In this section, reduced-order FIR models are determined using a least-squares cost function, and several limitations thereof are shown, thereby constituting \contributionRef{Contribution:MRKernel:i}. 
\else
	In this section, reduced-order FIR models are determined using a least-squares cost function, and several limitations thereof are shown, thereby constituting contribution C1. 
\fi
Specifically, these limitations include that the model order is restricted to the length of the data of the slow-rate output. Additionally, the input to the system cannot be zero-order hold, and hence, needs to be excited at the fast-rate.

First, the fast-rate model output in \eqref{MRKernel:eq:fastSampledOutputTime} is vectorized as
\begin{equation}
	\label{MRKernel:eq:LiftedFastSampledModelOutput}
	\begin{aligned}
		\hat{y}_h = \begin{bmatrix}
			\hat{y}_h(0) &
			\hat{y}_h(1) &
			\cdots &
			\hat{y}_h(N-1)
		\end{bmatrix}^\top.
	\end{aligned}
\end{equation}
Downsampling $\hat{y}_h$ in \eqref{MRKernel:eq:LiftedFastSampledModelOutput} with a factor $F$, which essentially involves selecting every $\fac^\mathrm{th}$ value of $\hat{y}_h$, results in the slow-rate model output $\hat{y}_l$, i.e., 
\begin{equation}
	\label{MRKernel:eq:slowSampledModelOutput}
	\begin{aligned}
		\hat{y}_l &\!=\!\! \begin{bmatrix}
			\hat{y}_l(0) &
			\hat{y}_l(1) &
			\cdots &
			\hat{y}_l(M-1)
		\end{bmatrix}^\top  = \Phi \theta \\
	\end{aligned}
\end{equation}
with parameter vector $\theta = \begin{bmatrix}
	\theta_0 & \theta_1 & \cdots & \theta_{P-1}
\end{bmatrix}^\top$ and regressor matrix $\Phi\in\mathbb{R}^{M\times P}$, that is
\begin{equation}
	\begin{aligned}
		\Phi = \begin{bmatrix}
			u_h(0) 					& 0 		& \cdots 		&\cdots				&0 \\
			u_h(\fac) 				&  \cdots 	&  u_h(0) 	&\cdots 			& 0 \\
			\vdots 					&  \vdots 	& \vdots 		& \vdots 				&  \vdots \\
			u_h\left(\left(M-1\right)\fac\right) 		& \cdots 	&\cdots 		&\cdots				 &  u_h\left(\left(M-1\right)\fac-P\right)
		\end{bmatrix}.
	\end{aligned}
\end{equation}

The goal is to identify $\theta$ given $u_h$ and $y_l$. Similar to identifiability of fast-rate models in \defRef{def:Identifiability}, unique solutions for $\theta$ given $u_h$ are defined as
	\begin{equation}
		\label{MRKernel:eq:UniqueFastRateModel}
		\begin{aligned}
			{S}_d\Psi\left(\theta-\theta^*\right)=\Phi\left(\theta-\theta^*\right)=0,\quad\Rightarrow \theta=\theta^*.
		\end{aligned}
	\end{equation}
Unique fast-rate models are analyzed in \lemmaRef{lemma:SlowSampledIdentifiability} and \lemmaRef{lemma:fastSampledInput}.
\begin{lemma}
	\label{lemma:SlowSampledIdentifiability}
	If $P\geq M$, there does not exist a unique fast-rate model $G(q,\theta)$ as defined in \eqref{MRKernel:eq:UniqueFastRateModel}.
\end{lemma}
\begin{proof}
Since matrix $\Phi\in\mathbb{R}^{M\times P}$ with $P>M$ has more columns than rows, the columns must be linearly dependent. This means there exists a non-zero vector $\left(\theta-\theta^*\right)\in\mathbb{R}^P$ such that $\Phi\left(\theta-\theta^*\right)=0$, resulting in a non-unique model according to \eqref{MRKernel:eq:UniqueFastRateModel}.
\end{proof}

\lemmaRef{lemma:SlowSampledIdentifiability} illustrates that no more parameters $\theta$ can be identified than the $M$ samples of the output $y_l$.
\begin{lemma}
	\label{lemma:fastSampledInput}
	There does not exist a unique fast-rate model $G(q,\theta)$ with $P>2$, given $u_h$ and $y_l$, for zero-order hold input $u_h(m\fac+i)=u_h(m\fac)=u_l(m)\: \forall i\in\mathbb{N}_{[0,\fac-1]}$.
\end{lemma}
\begin{proof}
The regressor matrix $\Phi\in\mathbb{R}^{M\times P}$ for zero-order hold inputs is equal to the first $P$ columns of the matrix
\begin{equation}
	\label{MRKernel:eq:compartPhi}
	\begin{aligned}
			\begin{bmatrix}
		u_l	& \Phi_u \otimes {1}_{1\times\fac} & {0}_{M\times\fac-1}
	\end{bmatrix}=	\left[\begin{array}{c|c}	\vphantom{\Phi^*}\Phi & \Phi^* \end{array}\right] \in\mathbb{R}^{M\times N},
	\end{aligned}
\end{equation}
where $\Phi^*\in\mathbb{R}^{M\times N-P}$ and
\begin{equation}
	\begin{aligned}
		\Phi_u =  \begin{bmatrix}
		0  & 0 & \cdots & 0\\
		u_l(0) & 0 & \cdots& \vdots\\
		u_l(1) & u_l(0) & \cdots & \vdots\\
		\vdots  & \vdots & \ddots & \vdots\\
		u_l\left(M- 2\right) & u_l\left(M- 3\right) & \cdots & u_l(0)
	\end{bmatrix} \in \mathbb{R}^{M\times M-1}.
	\end{aligned}
\end{equation}
As a result of the repeating columns of $\Phi$ in \eqref{MRKernel:eq:compartPhi}, the matrix $\Phi$ has non-zero null-space. Therefore, there does not exist a unique fast-rate model $G(q,\theta)$ with $P>2$ as defined in \eqref{MRKernel:eq:UniqueFastRateModel} for zero-order hold inputs.
\end{proof}

\lemmaRef{lemma:fastSampledInput} illustrates that zero-order hold inputs cannot be used to uniquely identify fast-rate models with $P>2$.

The above conditions have to be taken into account for obtaining a unique estimate from a least-squares cost function. For estimating fast-rate models using slow-rate output measurements, the least-squares problem minimizes the squared error between measured and estimated slow-rate outputs \eqref{MRKernel:eq:slowSampledModelOutput} as
\begin{equation}
	\label{MRKernel:eq:LSCostFIR}
	\begin{aligned}
		\min_{\theta} V(\theta) 
		=\min_{\theta} \left\|
		y_l-\Phi\theta
		\right\|_2^2.
	\end{aligned}
\end{equation}
The least-squares problem \eqref{MRKernel:eq:LSCostFIR} has minimizer
\begin{equation}
	\label{MRKernel:eq:FIRModelSolution}
	\begin{aligned}
		\theta = \left(\Phi^\top\Phi\right)^{-1}\Phi^\top y_l.
	\end{aligned}
\end{equation}
\lemmaRef{lemma:SlowSampledIdentifiability} illustrates that fast-rate FIR models cannot be uniquely determined for order $P> M$ using this generic least-squares cost function. Additionally, \lemmaRef{lemma:fastSampledInput} shows that fast-rate models cannot be uniquely determined for zero-order hold inputs, and therefore, fast-rate inputs are required. 
\section{Kernel-Regularized Identification Above the Nyquist Frequency}
\label{MRKernel:sec:KernelFIR}
\ifthesismode
	In this section, accurate fast-rate models beyond the Nyquist frequency of slow-rate output measurements are determined for arbitrary orders $P\in\mathbb{N}_{[1,N]}$, including non-parametric models $P=N$, hence relaxing the limitation of $P<M$ and constituting \contributionRef{Contribution:MRKernel:ii}.
\else
	In this section, accurate fast-rate models beyond the Nyquist frequency of slow-rate output measurements are determined for arbitrary orders $P\in\mathbb{N}_{[1,N]}$, including non-parametric models $P=N$, hence relaxing the limitation of $P<M$ and constituting contribution C2. 
\fi
This is realized through the use of regularization techniques that enforce certain properties of the estimated model.

First, the regularized estimation problem is posed. Second, several design aspects are illustrated, including regularization options and a procedure that summarizes the approach.
\subsection{Regularized FIR Estimation Above the Nyquist Frequency}
Properties of the estimated FIR coefficients $\theta$ are enforced by regularizing the cost function in \eqref{MRKernel:eq:LSCostFIR} with $\theta^\top K^{-1}\theta$, i.e.,
\begin{equation}
	\label{MRKernel:eq:LSKernelCostFIR}
	\begin{aligned}
		&\min_{\theta} V(\theta)+ \gamma \theta^\top K^{-1} \theta\\
		= &\min_{\theta} \left\|
		y_l-\Phi\theta
		\right\|_2^2 + \gamma \theta^\top K^{-1} \theta
	\end{aligned}
\end{equation}
with regularization parameter $\gamma$ and regularization or kernel matrix $K$. The optimal solution to \eqref{MRKernel:eq:LSKernelCostFIR} results in regularized least-squares estimation of fast-rate FIR models using slow-rate output measurements, and is given by
\begin{equation}
	\label{MRKernel:eq:FIRModelKernelSolution}
	\begin{aligned}
		\theta = K\Phi^\top\left(\Phi K \Phi^\top+\gamma I_M\right)^{-1}y_l,
	\end{aligned}
\end{equation}
with $\Phi$ from \eqref{MRKernel:eq:slowSampledModelOutput}, and regularization matrix 
\begin{equation}
	\label{MRKernel:eq:kernelMatrix}
	\begin{aligned}
		K=\begin{bmatrix}
			k(0,0) & k(0,1)	& \cdots & k(0,N-1) \\
			k(1,0)& k(1,1)	& \cdots & k(1,N-1) \\
			\vdots & \vdots & \ddots & \vdots \\
			k(N-1,0)& k(N-1,1)	& \cdots & k(N-1,N-1) \\
		\end{bmatrix}.
	\end{aligned}
\end{equation}
The entries $k(i,j)$ of regularization matrix $K$ enforce certain high-level properties on the FIR coefficients, and are treated in more detail in \secRef{MRKernel:sec:kernelDesign}. The fast-rate FIR model is uniquely determined as shown in \theoremRef{theorem:GlobalIdentifiabilityFIRRegularized}, leading to the main result in this section.
\begin{theorem}
	\label{theorem:GlobalIdentifiabilityFIRRegularized}
	The fast-rate FIR model using slow-sampled outputs of order $P\in\mathbb{N}_{[1,N]}$, when determined using the regularized solution in \eqref{MRKernel:eq:FIRModelKernelSolution}, is uniquely determined if $\gamma>0$.
\end{theorem}
\begin{proof}
	$\Phi K \Phi^\top+\gamma I_M$ is invertible when $\gamma>0$, since 
	\begin{equation}
		\begin{aligned}
			x^\top \left(\Phi K \Phi^\top+\gamma I_M\right) x &= x^\top\Phi K \Phi^\top x+\gamma x^\top x > 0, \\
			 &\forall x\in\mathbb{R}^{M} \setminus \{ 0 \}. 
		\end{aligned}
	\end{equation}
	Therefore, $K\Phi^\top\left(\Phi K \Phi^\top+\gamma I_M\right)^{-1}y_l$ from \eqref{MRKernel:eq:FIRModelKernelSolution} has a unique solution when $\gamma>0$.
\end{proof}

In sharp contrast to the reduced-order least-squares estimator, the regularized least-squares estimator uniquely determines fast-rate FIR models of order $P\in\mathbb{N}_{[1,N]}$, independent of the order $P$ or fast-rate inputs $u_h$. This is explained because the regularized least-squares estimator incorporates certain prior information about the fast-rate impulse response coefficients by using the regularization matrix $P$ as the prior covariance matrix of $\theta$ \citep{Pillonetto2014}. An intuitive example illustrating that the regularized least-squares estimator is always capable of estimating a model of arbitrary order is given in \exampleRef{example:lowOrderP}.
\begin{example}
	\label{example:lowOrderP}
	Let the length of the data in the slow-rate output be $M=1$, i.e., $y_l\in\mathbb{R}$. If $\gamma>0$, the model \eqref{MRKernel:eq:FIRModelKernelSolution} is uniquely determined for order $P\in\mathbb{N}_{[1,N]}$ for any $N>1$, according to \theoremRef{theorem:GlobalIdentifiabilityFIRRegularized}. In sharp contrast, a fast-rate model determined with unregularized least-squares \eqref{MRKernel:eq:FIRModelSolution} can only be uniquely identified for $P=1$.
\end{example}

\exampleRef{example:lowOrderP} illustrates that the regularized least-squares estimator is capable of estimating a fast-rate model of arbitrary order even when the slow-rate output measurements are not informative. This is explained because the kernel-regularized estimator will always return a model of arbitrary order, independent of data length or persistence of excitation, by specifying prior information through the regularization matrix $P$.

\begin{remark}
		Note that there is an equivalence between kernel regularized estimates and Gaussian process regression \citet[Section~4.3]{Pillonetto2014}. In addition, Gaussian process prediction, or reformulating (15) as a reproducing kernel Hilbert space function estimation problem \citep[Section~9]{Pillonetto2014}, enables estimating the FIR model at arbitrary time indices.
\end{remark}

\subsection{Design Aspects}
\label{MRKernel:sec:kernelDesign}
In this section, several kernel functions are introduced and the developed approach is summarized in a procedure.

The selection of kernel functions $k(i,j)$ is essential, since they determine key regularization properties of the estimated fast-rate FIR model, including smoothness, decay, and resonant dynamics. Different choices of kernel functions impose different structural constraints, which influence a trade-off between model flexibility and model fit. The most generic kernel matrix is the identity matrix, i.e.,
\begin{equation}
	\label{MRKernel:eq:TikhonovKernel}
	\begin{aligned}
		k_{T}(i,j) = \begin{cases}
			1,&i=j, \\
			0, & i\neq j,
		\end{cases}
	\end{aligned}
\end{equation}
that is generally referred to as ridge regression or Tikhonov regularization \citep{Tikhonov1977}, that enables estimating $P>M$ FIR coefficients, in exchange for a certain amount of bias. Alternatively, kernels encoding properties that are reasonable to assume for impulse responses include the Diagonal-Correlated (DC) and stable-spline kernels. The DC kernel is given by
\begin{equation}
	\label{MRKernel:eq:DC}
	\begin{aligned}
		k_{DC}(i,j) = \lambda \alpha^{\left(i+j\right)/2}\beta^{|j-i|},
	\end{aligned}
\end{equation}
where $0\leq\alpha<1$ accounts for exponential decay and $|\beta|<1$ describes the correlation between neighboring impulse response coefficients. The stable-spline kernel is given by
\begin{equation}
	\label{MRKernel:eq:SS}
	\begin{aligned}
		k_{SS}(i,j) = \lambda \left(\frac{\alpha^{i+j+\max (i, j)}}{2}-\frac{\alpha^{3 \max (i, j)}}{6}\right),
	\end{aligned}
\end{equation}
where $\alpha$ plays a similar role as for the DC kernel. Finally, the prior knowledge kernel, which encodes information about poles of a system, is given by \citep{Hallemans2022}
\begin{equation}
	\label{MRKernel:eq:priorKnowledgeKernel}
	\begin{aligned}
		k_{pk,n}(i,j) = \alpha_n^{\left(i+j\right)/2}\bigg(\gamma_{1,n}\cos\left(\omega_n\left(i- j\right)\right)+\gamma_{2,n}\cos\left(\omega_n \left(i+j\right)\right)\bigg)
	\end{aligned}
\end{equation}
where $\alpha_n$ has a similar role as for the DC kernel \eqref{MRKernel:eq:DC}, $\omega_n \in \mathbb{R}_{[0,2\pi)}$ is the normalized frequency of the pole, and
\begin{equation}
	\label{MRKernel:eq:ZetaGamma}
	\begin{aligned}
		\gamma_{1, n}=\frac{\sigma_{1, n}^2+\sigma_{2, n}^2}{2}, && \gamma_{2, n}=\frac{\sigma_{1, n}^2-\sigma_{2, n}^2}{2},
	\end{aligned}
\end{equation}
where $\sigma_{1,n}^2$ and $\sigma_{2,n}^2$ are the variances related to the response of the pole from the prior knowledge kernel \citep{Hallemans2022}. The dependence on $n$ allows the prior knowledge kernel to encode information about multiple poles, which is achieved by summing the kernel functions over $n$. The prior knowledge kernel is especially effective in estimating systems with lightly-damped resonant dynamics, which can generally not be estimated accurately using kernels that assume a decaying impulse response such as the DC and stable-spline kernels.

\begin{remark}
	Note that for $\gamma=0$, the developed kernel-regularized method \eqref{MRKernel:eq:FIRModelKernelSolution} recovers the least-squares FIR estimator \eqref{MRKernel:eq:FIRModelSolution} described in \secRef{MRKernel:sec:reducedFIR}.
\end{remark}

The kernel functions depend on the parameters $\lambda$, $\alpha$, $\rho$, $\alpha_n$, $\omega_n$, and $\gamma_{i,n}$, which are typically referred to as hyperparameters. 
Selecting the hyperparameters is crucial, since they directly influence high-level properties of the fast-rate FIR model, such as the exponential decay $\alpha$ and the pole frequency $\omega_n$.

The hyperparameters can be tuned in several ways. First, the hyperparameters can be optimized using the marginal likelihood \citep{Pillonetto2014}
\begin{equation}
	\label{MRKernel:eq:MarginalLikelihood}
	\begin{aligned}
		\hat{\eta} = \arg\min_\eta y_l^\top \left(\Phi K(\eta)\Phi^\top+\gamma I_M\right)^{-1}y_l+\log \det \left(
		\Phi K(\eta)\Phi^\top\right),
	\end{aligned}
\end{equation}
where the kernel matrix $K(\eta)$ from \eqref{MRKernel:eq:kernelMatrix} is parameterized using the hyperparameters $\eta$. Additionally, the regularization parameter $\gamma$ can be included in $\eta$ and optimized simultaneously. Since only slow-rate output measurements are available, marginal-likelihood \eqref{MRKernel:eq:MarginalLikelihood} optimizes the hyperparameters based on their effect on the slow-rate output measurements $y_l$. As a result, marginal likelihood optimization effectively selects the hyperparameters that balances slow-rate output data fit and model complexity \citep{Rasmussen2004}.

Alternatively, an additional data set can be used for cross-validation to optimize the hyperparameters by selecting them to generalize well to unseen data \citep{Pillonetto2014}, which is shown to be effective in practice.

The kernel functions encode prior information about the impulse response coefficients, and hence, introduce correlation between the coefficients that enable estimating the $N$ coefficients from $M<N$ samples in $y_l$.

The developed approach is summarized in Procedure~\ref{MRKernel:proc:1}, that links the main results in this \manuscript.
\begin{figure}[tb]
	\vspace{-10pt}\hrule \vspace{1mm}\begin{proced}[Identify fast-rate FIR model of arbitrary order using slow-rate output measurements and fast-rate inputs] \hfill \vspace{0.5mm} \hrule \vspace{1mm}
		\label{MRKernel:proc:1}
		\begin{enumerate}
			\item Construct fast-rate input $u_h$, e.g., random noise or a random binary sequence.
			\item Apply input $u_h$ to the system and record $y_l$.
			\item Construct the regressor matrix $\Phi$ according to \eqref{MRKernel:eq:slowSampledModelOutput}.
			\item (Optional:) Tune the hyperparameters $\eta$ and $\gamma$, e.g., based on marginal likelihood optimization \eqref{MRKernel:eq:MarginalLikelihood}.
			\item Construct kernel matrix $K$ using \eqref{MRKernel:eq:kernelMatrix}.
			\item Calculate the $P$ FIR coefficients $\theta$ via least-squares \eqref{MRKernel:eq:FIRModelSolution} and kernel-regularized \eqref{MRKernel:eq:FIRModelKernelSolution} approaches.
		\end{enumerate}
		\vspace{0pt} 	\hrule \vspace{-9pt}
	\end{proced}
\end{figure}
\section{Simulation Example}
\label{MRKernel:sec:simulations}
\ifthesismode
	In this section, the developed method is validated using a Monte Carlo simulation, contributing to \contributionRef{Contribution:MRKernel:iii}.
\else
	In this section, the developed method is validated using a Monte Carlo simulation, contributing to contribution C3.
\fi
\subsection{Simulation Setup}
\label{MRKernel:sec:simSetup}
\begin{figure}[tb]
	\centering
	\includegraphics{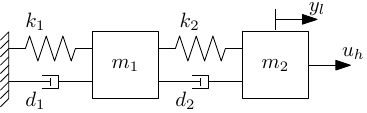}
	\caption{Mass-spring-damper system considered.}
	\label{MRKernel:fig:systemExample}
\end{figure}

\begin{figure}[tb]
	\centering
	\includegraphics{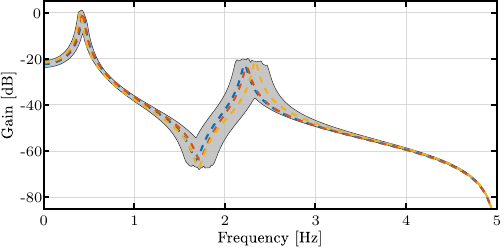}
	\caption{The FRFs during 100 Monte Carlo simulations are contained in \grayLegend, where three FRFs are shown as (\markerlineNoBracket{mblue}[densely dashed], \markerlineNoBracket{mred}[densely dashed], \markerlineNoBracket{myel}[densely dashed]).}
	\label{MRKernel:fig:SimFRFs}
\end{figure}

A Monte-Carlo simulation study is carried out for the mass-spring-damper system in \figRef{MRKernel:fig:systemExample}, where the settings are listed in \tabRef{MRKernel:tab:simValues}.
The nominal system parameters are given by 
\begin{equation*}
	\begin{aligned}
		k_1&=15\text{ N/m}, & k_2&=100\text{ N/m}, & d_1 &= 0.45\text{ Ns/m}, \\
		d_2&=0.06\text{ Ns/m}, & m_1&=1\text{ kg}, & m_2 &= 1 \text{ kg}.
	\end{aligned}
\end{equation*}
During every Monte Carlo iteration, each parameter is perturbed randomly within 10\% of its nominal value. The FRFs of the system during the Monte Carlo runs are seen in \figRef{MRKernel:fig:SimFRFs}.
Additional simulation settings are seen in \tabRef{MRKernel:tab:simValues}.
\begin{table}[tb]
	\centering
	\caption{Simulation settings.}
	\label{MRKernel:tab:simValues}
	\begin{tabular}{llll}
		\toprule
		\textbf{Variable}    & \textbf{Abbrevation} & \textbf{Value} & \textbf{Unit} \\
		\midrule
		Fast sampling time   & \tsh            & 0.1       & s   \\
		Slow sampling time & \tsl 				& 0.3 		& s \\
		Downsampling factor  & \fac            & 3       & -    \\
		Number of input samples    & $N$               & 600    & -    \\
		Number of output samples    & $M$               & 200    & -    \\
		Monte Carlo runs	& -					& 100	& -	\\
		\bottomrule
	\end{tabular}
\end{table}

Independent training and validation data sets are generated, where the excitation signals $u_h$ are random phase multisines with the same flat amplitude spectrum but different phase realizations. The output of the training set is disturbed with additive white Gaussian noise $e_h$, where the signal-to-noise ratio, that is the ratio between the variance of $y_h$ and $e_h$, is randomly chosen in $\mathbb{R}_{[40,\: 60]}$. 

The following models are estimated using training data and compared on validation data in each Monte Carlo run.
\begin{itemize}
	\item[$G_{LS}$] Unregularized least-squares FIR from \eqref{MRKernel:eq:FIRModelSolution} of order $P\in\mathbb{N}_{[50,N]}$.
	\item[$G_{DC}$] Kernel regularized FIR using \eqref{MRKernel:eq:FIRModelKernelSolution} of order $P\in\mathbb{N}_{[50,N]}$, which uses the DC kernel $k_{DC}$ \eqref{MRKernel:eq:DC}.
	\item[$G_{PK}$] Kernel regularized FIR using \eqref{MRKernel:eq:FIRModelKernelSolution} of order $P\in\mathbb{N}_{[50,N]}$, which uses the kernel function with prior knowledge from \eqref{MRKernel:eq:priorKnowledgeKernel}, specifically
	\begin{equation}
		\label{MRKernel:eq:KPKSim}
		\begin{aligned}
			k_{PK}(i,j)= k_{DC}(i,j)+\sum_{n=1}^{2} k_{pk,n}(i,j).
		\end{aligned}
	\end{equation}
\end{itemize}

The hyperparameters of the kernel regularized estimators are chosen as 
\begin{equation*}
	\begin{aligned}
		\lambda&=1, & 	\omega_1&=0.4\tsh\cdot2\pi, & \omega_2 &= 2\tsh\cdot2\pi, \\
		 \alpha&=e^{-0.5 \tsh}, & \alpha_1&=\alpha_2 = e^{-0.5 \tsh}, & \gamma&=10^{-5},\\
		  \beta&=e^{-0.1 \tsh}, &  \sigma_{1,n}&=\sigma_{2,n} = 1,\\
	\end{aligned}
\end{equation*} 
where $\gamma_{1, n}$ and $\gamma_{2, n}$ are determined using \eqref{MRKernel:eq:ZetaGamma}. Additionally, the regularization parameter $\gamma$ and the hyperparameters $\omega_n$ and $\alpha_n$ are optimized using marginal likelihood optimization \eqref{MRKernel:eq:MarginalLikelihood} with a gradient-based constrained optimizer to further enhance accuracy.
\subsection{Simulation Results}
\begin{figure}[tb]
	\centering
	\includegraphics{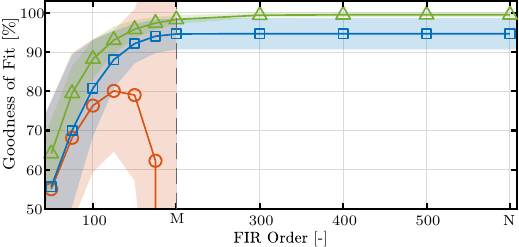}
	\caption{Kernel regularized estimators using DC kernel \eqref{MRKernel:eq:DC} \markerline{mblue}[solid][square][2][0.8] or prior knowledge kernel \eqref{MRKernel:eq:KPKSim} \markerline{mgreen}[solid][triangle][2][0.8] are capable of achieving a higher GoF for all model orders compared to least-squares FIR models \markerline{mred}[solid][o][2][0.8]. In addition, the mean GoF $\pm 2$ times its sample standard deviation (shaded areas) show that the prior knowledge kernel decreases variance.}
	\label{MRKernel:fig:GOFSim}
\end{figure}

\begin{figure}[tb]
	\centering
	\includegraphics{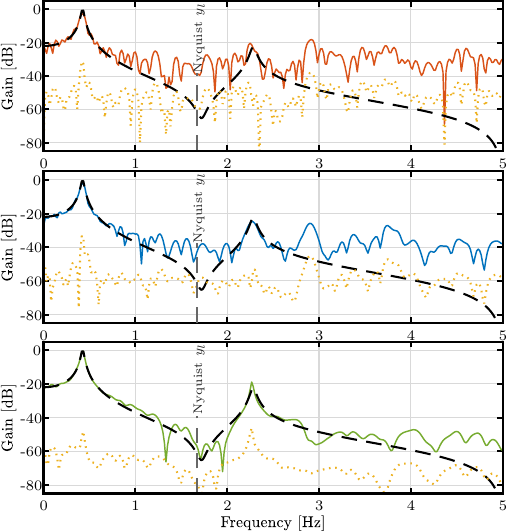}
	\caption{
		The regularized FIR models with order $P=N=600$ improve the identification of the true FRF \markerline{black}[densely dashed] using a DC kernel \markerline{mblue} (middle) and a prior knowledge kernel \markerline{mgreen} (bottom). The least-squares FIR model with highest GoF having order $P=150$ \markerline{mred} (top) captures only the first resonance. The regularized FIR model with DC kernel partially captures the second resonance, while the prior knowledge kernel model captures both resonances. These differences are reflected in the scaled FRF error $0.1\cdot\left|G_0\left(e^{j\omega_k\tsh}\right)-G\left(e^{j\omega_k\tsh}\right)\right|$ \markerline{myel}[densely dotted], which is lower for the regularized models.
	}
	\label{MRKernel:fig:SimFRF}
\end{figure}

For each Monte Carlo run, the Goodness of Fit (GoF) of the fast-rate output $y_h$ on the validation data is calculated,
\begin{equation}
	\label{MRKernel:eq:GOF}
	\begin{aligned}
		\text{GoF} = 100\left(1-\frac{\sum_{\dt=0}^{N-1}\left(y_{h,v}(\dt)-\hat{y}_h(\dt)\right)^2}{\sum_{\dt=0}^{N-1}\left(y_{h,v}(\dt)-\bar{y}_{h,v}\right)^2}\right),
	\end{aligned}
\end{equation}
where $\hat{y}_h$ is estimated using the fast-rate FIR models, $y_{h,v}(\dt)$ are the validation values, and $\bar{y}_{h,v}$ is the mean value of $y_{h,v}(\dt)$. The mean plus and minus 2 times its sample standard deviation over the Monte Carlo runs are seen in \figRef{MRKernel:fig:GOFSim}.
Additionally, the fast-rate FRFs have been evaluated for one Monte Carlo run, and are seen in \figRef{MRKernel:fig:SimFRF}.
The following observations are made from the simulation results.
\begin{itemize}
	\item From the GoF in \figRef{MRKernel:fig:GOFSim}, the following is observed.
	\begin{itemize}
		\item The least-squares FIR model is capable of estimating a model of order $P<M$, but not for $P>M$, supporting \lemmaRef{lemma:SlowSampledIdentifiability}.
		\item The kernel regularized FIR model with DC kernel accurately identifies models with low variance for all $P\in\mathbb{N}_{[50,N]}$, due to the additional prior that is encoded in the estimator.
		\item The additional prior information about the system's poles introduced by the prior-knowledge kernel \eqref{MRKernel:eq:KPKSim} enables a further improvement in GoF and a reduction in variance for $P>200$, achieving a GoF of 99.50\% compared to 94.67\% for the DC kernel.
	\end{itemize}
	\item From the FRFs in \figRef{MRKernel:fig:SimFRF}, the following is observed.
	\begin{itemize}
		\item The FRF of the least-squares FIR model with $P=150$ in the top doesn't identify the FRF accurately, especially beyond the Nyquist frequency.
		\item The FRF of the regularized FIR models in the middle and bottom are significantly more accurate with lower FRF error, especially beyond the Nyquist frequency of the slow-rate output. In particular, the FIR model with prior knowledge kernel (bottom) accurately identifies the resonant behavior.
	\end{itemize}
\end{itemize}
\section{Experimental Validation}
\label{MRKernel:sec:exp}
In this section, the developed method is validated using an experimental prototype wafer stage with a slow-rate output, contributing to contribution C3.
\subsection{Experimental Setup}
\begin{figure}[tb]
	\centering
	\setlength{\fboxsep}{0pt}%
	\setlength{\fboxrule}{0.75pt}%
	\begin{subfigure}[tb]{\linewidth}
		\centering
		\includegraphics{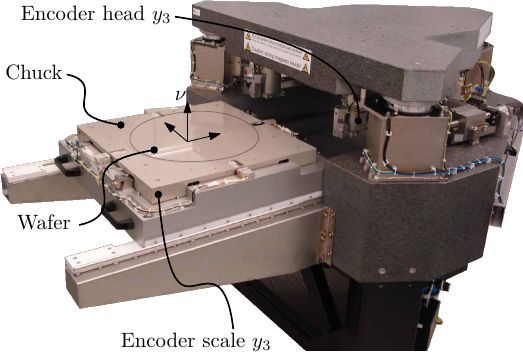}
		\caption{Photograph of experimental setup.}
		\label{MRKernel:fig:photoOAT}
	\end{subfigure}
	\ifthesismode
		\vspace{10mm}
	\fi
	\begin{subfigure}[tb]{\linewidth}
		\centering
		\includegraphics{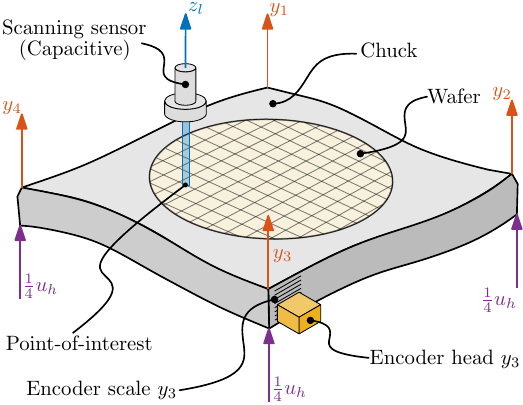}
		\caption{Schematic overview of experimental setup.}
		\label{MRKernel:fig:schematicOAT}
	\end{subfigure}
	\caption{Prototype experimental wafer stage setup used in semiconductor manufacturing with a slow-rate output.}
	\label{MRKernel:fig:OAT}
\end{figure}

The experimental setup is the prototype wafer stage in \figRef{MRKernel:fig:OAT} \citep{Classens2023}, which is used in semiconductor manufacturing and a prime example of mechatronic systems with slow-rate outputs. 

The objective of the wafer stage is to accurately control the vertical displacement $\nu$ of the point-of-interest on the wafer shown in \figRef{MRKernel:fig:photoOAT}, which is the point on the wafer during lithographic exposure. The chuck and wafer are levitated, and actuated by four Lorentz actuators. The vertical displacement of the chuck is measured by four linear encoders and used as inputs to the feedback controller as $y=\frac{1}{4}\left(y_1+y_2+y_3+y_4\right)$. Directly measuring the vertical displacement of the point-of-interest on the wafer $\nu$ is not possible using linear encoders. The chuck of the wafer stage has internal lightly-damped structural modes. Consequently, measuring the vertical displacement on the outside of the chuck does not coincide with the vertical displacement at the wafer's point of interest \citep{Oomen2015,Heertjes2020}. Therefore, an external capacitive sensor, denoted by scanning sensor in \figRef{MRKernel:fig:schematicOAT}, directly measures the vertical displacement of the point-of-interest. The external capacitive sensor is sampled at a reduced sampling rate compared to the actuators, resulting in a slow-rate output.
\ifthesismode
\FloatBarrier
\fi
\subsection{Identification Setting}
\begin{figure}[tb]
	\centering
	\includegraphics{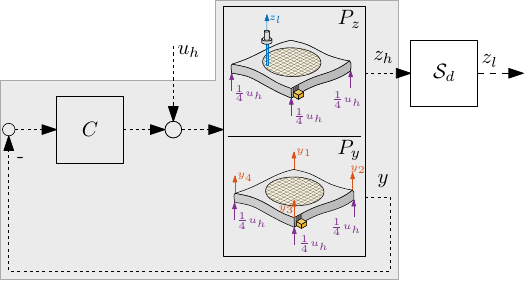}
	\caption{Experimental feedback scheme used, where the equivalent system in \eqref{MRKernel:eq:EquivalentSystem} is to be identified.}
	\label{MRKernel:fig:ExperimentalScheme}
\end{figure}

For this experimental validation, the goal is to identify the fast-rate equivalent model of the prototype wafer stage
\begin{equation}
	\label{MRKernel:eq:EquivalentSystem}
	\begin{aligned}
		G_z\left( q,\theta\right)  = P_z\left(\vphantom{P_y\left( q,\theta\right)} q,\theta\right) \left(I+C\left( q\right) P_y\left( q,\theta\right) \right)^{-1}: u_h\mapsto z_h,
	\end{aligned}
\end{equation}
using the fast-rate excitation $u_h$ and slow-rate outputs $z_l$. The prototype wafer stage is operating in closed-loop control, where the control structure is seen in \figRef{MRKernel:fig:ExperimentalScheme}.
For this experimental validation, the fast-rate excitation signal $u_h$ is equally distributed over the four actuators, and is considered a disturbance to the plant. The excitation signal $u_h$ is a random noise sequence with flat amplitude spectrum, resulting in a signal-to-noise ratio of approximately 45 dB. The scanning sensor is suspended above the wafer, and is positioned in the bottom left corner of the wafer, 110 mm in both directions from the center. The experimental settings are seen in \tabRef{MRKernel:tab:expValues}.
\begin{table}[tb]
	\centering
	\caption{Experimental settings.}
	\label{MRKernel:tab:expValues}
	\begin{tabular}{llll}
		\toprule
		\textbf{Variable}    & \textbf{Abbr.} & \textbf{Value} & \textbf{Unit} \\
		\midrule
		Fast sampling time   & \tsh            & 0.5       & ms   \\
		Slow sampling time & \tsl 				& 1.5 		& ms \\
		Downsampling factor  & \fac            & 3       & -    \\
		Number of input samples    & N               & 600    & -    \\
		Number of output samples    & M               & 200    & -    \\
		\bottomrule
	\end{tabular}
\end{table}
Similar to \secRef{MRKernel:sec:simulations}, FIR models are identified using unregularized and regularized least-squares cost functions, i.e., the following methods are compared.
\begin{itemize}
	\item[$G_{LS}$] Unregularized least-squares FIR from \eqref{MRKernel:eq:FIRModelSolution} of order $P\in\mathbb{N}_{[5,N]}$.
	\item[$G_{DC}$] Kernel regularized FIR using \eqref{MRKernel:eq:FIRModelKernelSolution} of order $P\in\mathbb{N}_{[5,N]}$, which uses the DC kernel $k_{DC}$ \eqref{MRKernel:eq:DC}.
	\item[$G_{PK}$] Kernel regularized FIR using \eqref{MRKernel:eq:FIRModelKernelSolution} of order $P\in\mathbb{N}_{[5,N]}$, which uses the kernel function with prior knowledge from \eqref{MRKernel:eq:priorKnowledgeKernel}, specifically
	\begin{equation}
		\label{MRKernel:eq:KPKExp}
		\begin{aligned}
			k_{PK}(i,j)= k_{DC}(i,j)+\sum_{n=1}^{3} k_{pk,n}(i,j).
		\end{aligned}
	\end{equation}
\end{itemize}
The hyperparameter of the kernels are initialized as
\begin{equation*}
	\begin{aligned}
		\lambda&=1, & \alpha &=\alpha_n= e^{-80 \tsh},\\
		 \omega_1&=490\tsh \cdot\!2\pi & \omega_2 &= 510\tsh\cdot\!2\pi ,&	\omega_3 &= 870\tsh\cdot\!2\pi,\\
		\beta&=e^{-20 \tsh}, &  \sigma_{1,n}&=\sigma_{2,n} = 1,  & \gamma&=2\cdot10^{-1}\\
	\end{aligned}
\end{equation*}
where $\gamma_{1,n}$ and $\gamma_{2,n}$ are determined using \eqref{MRKernel:eq:ZetaGamma}. The hyperparameters $\omega_n$ and $\alpha_n$ are refined using a gradient based constrained optimizer for the marginal likelihood \eqref{MRKernel:eq:MarginalLikelihood}.
\begin{figure}[tb]
	\centering
	\includegraphics{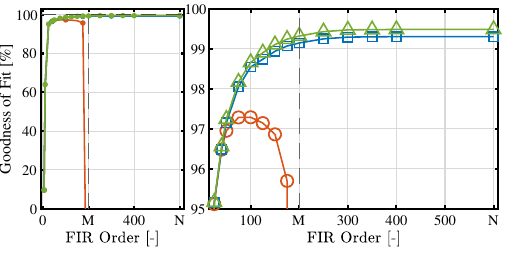}
	\caption{The GoF of the regularized FIR models with DC kernel \eqref{MRKernel:eq:DC} \markerline{mblue}[solid][square][1.5][0.7] and prior knowledge kernel \eqref{MRKernel:eq:KPKExp} \markerline{mgreen}[solid][triangle][2.5][0.7] are higher compared to the unregularized least-squares FIR model \markerline{mred}[solid][o][2][0.7] for all model orders $P\in\mathbb{N}_{[5,N]}$ (left) and $P\in\mathbb{N}_{[25,N]}$ (right).}
	\label{MRKernel:fig:ExpGoF}
\end{figure}

\ifthesismode
	\begin{figure}[tb]
\else
	\begin{figure*}[tb]
\fi
	\centering
	\ifthesismode
		\begin{subfigure}[t]{0.475\linewidth}
	\else
		\begin{subfigure}[b]{0.495\linewidth}
	\fi
		\centering
		\ifthesismode
			\includegraphics{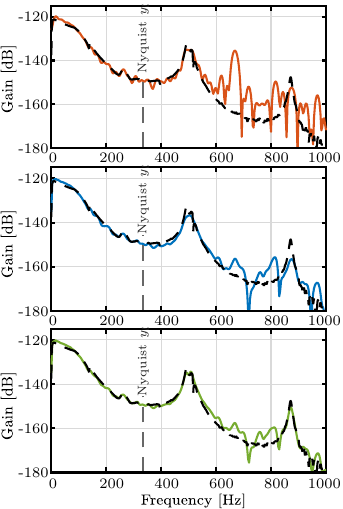}
		\else
			\includegraphics{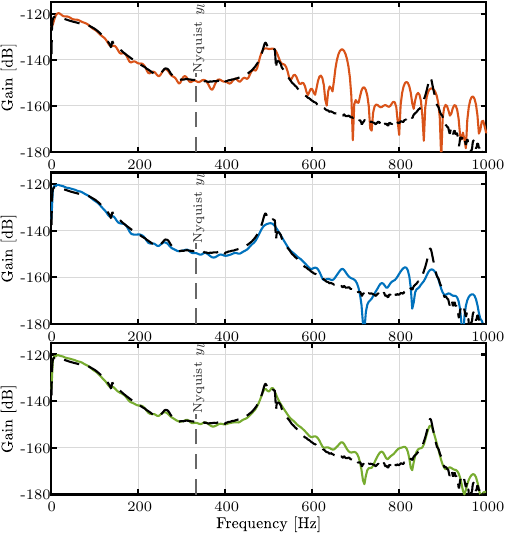}
		\fi
		\caption{The regularized FIR models with DC kernel \markerline{mblue} and prior knowledge kernel \markerline{mgreen} improve the estimate of the validation FRF $G_0\left(e^{j\omega_k\tsh}\right)$ \eqref{MRKernel:eq:ExpValidationFRF} \markerline{black}[densely dashed] compared to the least-squares FIR model \markerline{mred}.}
		\label{MRKernel:fig:ExpFRF}
	\end{subfigure} \hfill
	\ifthesismode
		\begin{subfigure}[t]{0.51\linewidth}
	\else
		\begin{subfigure}[b]{0.495\linewidth}
	\fi
		\centering
		\ifthesismode
			\includegraphics{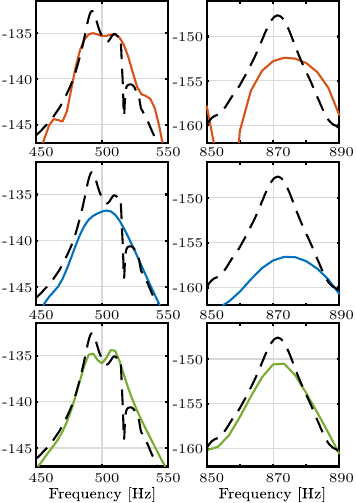}
		\else
			\includegraphics{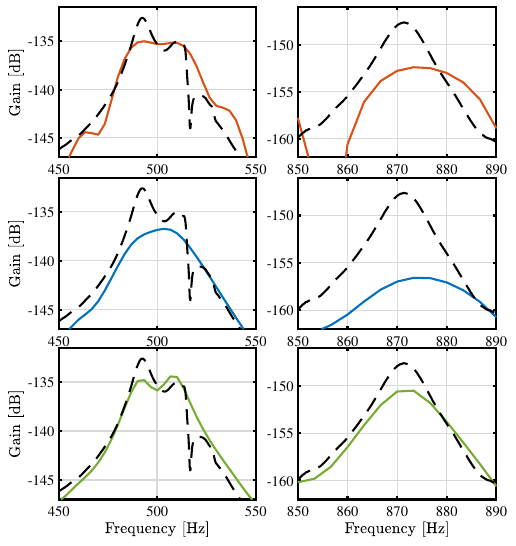}
		\fi
		\caption{Enlarged sections of the resonant behavior around 500 Hz (left) and 870 Hz (right) show that the FIR model with prior knowledge kernel accurately identifies the resonant behavior compared to the validation FRF $G_0\left(e^{j\omega_k\tsh}\right)$ \eqref{MRKernel:eq:ExpValidationFRF}.}
		\label{MRKernel:fig:ExpFRF_zoom}
	\end{subfigure}
	\caption{The regularized FIR models with order $P=N=600$ improve identification of the validation FRF $G_0\left(e^{j\omega_k\tsh}\right)$ \eqref{MRKernel:eq:ExpValidationFRF} for the prototype wafer stage \markerline{black}[densely dashed], using a DC kernel \markerline{mblue} (middle) and a prior knowledge kernel \markerline{mgreen} (bottom), compared to the least-squares FIR model with higest GoF having order $P=100<M$ \markerline{mred} (top).}
	\label{MRKernel:fig:ExpFRFTotal}
\ifthesismode
\end{figure}
\else
\end{figure*}
\fi

For validation purposes, the fast-rate outputs $z_h$ are recorded as well, and a validation data set with similar signal-to-noise ratio and the same amount of samples $N=600$ is measured to compare the approaches. 
Finally, a significantly longer data set containing $N=50000$ samples is measured, where $z_h$ is recorded as well, to identify the validation benchmark FRF
\begin{equation}
	\label{MRKernel:eq:ExpValidationFRF}
	\begin{aligned}
		G_0\left(e^{j\omega_k \tsh}\right): \frac{U_h\left(e^{j\omega_k \tsh}\right)}{\vphantom{U_h\left(e^{j\omega_k \tsh}\right)}Z_h\left(e^{j\omega_k \tsh}\right)}.
	\end{aligned}
\end{equation}
This validation FRF is identified with the local rational modeling method from \citet{McKelvey2012}, using local rational degrees $R=4$ and local window size $n_w = 150$.
\subsection{Experimental Results}
\begin{figure}[tb]
	\centering
	\ifthesismode
	\includegraphics{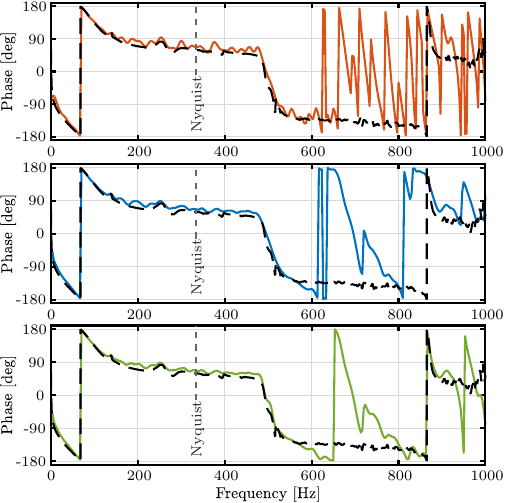}
	\else
	\includegraphics{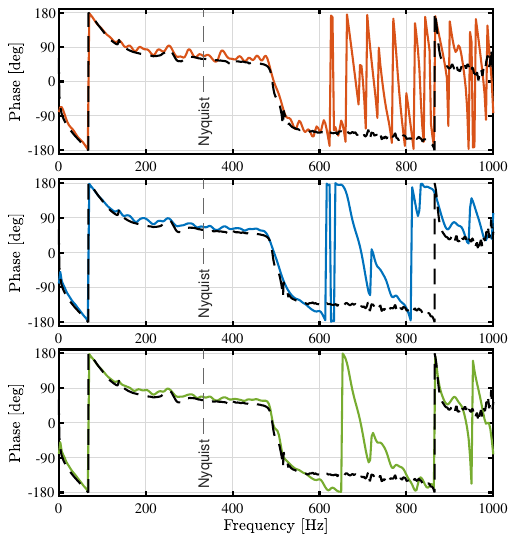}
	\fi
	\caption{Phase of prototype wafer stage \markerline{black}[densely dashed] estimated by unregularized FIR model \markerline{mred}, and regularized FIR models with DC kernel \markerline{mblue} and prior-knowledge kernel \markerline{mgreen}.}
	\label{MRKernel:fig:ExperimentalPhaseFIR}
\end{figure}

The GoF in \eqref{MRKernel:eq:GOF} for the validation data set is shown for various model orders in \figRef{MRKernel:fig:ExpGoF}.
The estimated FRFs are seen in \figRef{MRKernel:fig:ExpFRFTotal}, including an enlarged section of the resonant behavior around 500 and 870 Hz.
The estimated phase is shown in \figRef{MRKernel:fig:ExperimentalPhaseFIR}.
The following observations are made with respect to the experimental results.
\begin{itemize}
	\item From the GoF in \figRef{MRKernel:fig:ExpGoF}, the following is observed.
	\begin{itemize}
		\item The GoF for the regularized FIR model estimates using either the DC or prior knowledge kernel is high for all model orders, close to 100\% for $P\to N$. 
		\item The GoF for the least-squares FIR model estimates is significantly lower, and is not capable of estimating a model for $P>M$.
		\item The regularized FIR model with prior knowledge kernel and $P=N$ achieves a GoF of 99.44\%, and a root-mean-squared error of \[\sqrt{\frac{1}{N}\sum_{\dt=0}^{N-1}\left(y_{h,\nu}(\dt)-\hat{y}_h(\dt)\right)} = 1.65 \times 10^{-8} \text{ m},\] where the maximum absolute value and the RMS value of $y_{h,v}$ are $7.2\cdot10^{-7}$ m and $2.2\cdot10^{-7}$ m. In comparison, the least-squares FIR model with highest GoF 97.29\% has a root-mean-squared error of $3.62 \times 10^{-8}$ m, which is more than double.
	\end{itemize}
	\item From the FRFs in \figRef{MRKernel:fig:ExpFRFTotal}, the following is observed.
	\begin{itemize}
		\item Both the regularized FIR models with DC kernel and prior knowledge kernel estimate the FRF accurately, without erratic variations, even beyond the Nyquist frequency of the slow-rate output.
		\item The resonant behavior at 500 and 870 Hz in \figRef{MRKernel:fig:ExpFRF_zoom} is estimated accurately by the regularized FIR model with prior knowledge kernel, which is explained by the additional prior information that is introduced.
		\item The least-squares FIR model inaccurately estimates the FRF beyond 550 Hz. Furthermore, it introduces several artificial resonances, especially around $\fsl=666\frac{2}{3}$ Hz, which is caused by aliasing of the gain around 0 Hz to this frequency.
	\end{itemize}
\item \figRef{MRKernel:fig:ExperimentalPhaseFIR} shows that the phase of the wafer stage is estimated accurately for all methods up to 600 Hz. At higher frequencies, the regularized FIR model with prior-knowledge kernel provides the most accurate phase estimation.
\end{itemize}
\ifthesismode
\FloatBarrier
\fi

\section{Conclusions}
The results in this \manuscript determine conditions for unique identification of reduced-order FIR models and arbitrary order fast-rate models for systems with slow-rate output measurements, including non-parametric models, and additionally enables intersample response estimation. The method incorporates generic prior information through kernel regularization, that determines arbitrary order fast-rate models. This enables the usage of prior knowledge kernels that encode information about the poles of a system, resulting in a method which is highly suitable for lightly-damped systems.

The framework is validated using a Monte Carlo simulation example, which shows that the regularized FIR models have increased estimation quality and decreased variance. Furthermore, the framework is validated on a prototype wafer stage with a slow-rate output, showing that the regularized FIR models are capable of estimating the system, including lightly-damped resonant dynamics. The developed approach is a key enabler for control design of systems with slow-rate output measurements.

\section*{Acknowledgement}
The authors would like to thank Leonid Mirkin for his help and fruitful discussions, and Koen Classens for gathering experimental data, which both led to the results in this paper.
\bibliographystyle{elsarticle-num}
\bibliography{../../library}


\end{document}